\documentclass[aps,pre,reprint,amsmath,amssymb,groupedaddress,lengthcheck,showpacs]{revtex4-1}

\usepackage{graphicx} 
\usepackage{dcolumn} 
\usepackage{bm} 
\usepackage[hypertex]{hyperref}
\usepackage{color}
\usepackage{enumerate}

\begin{document}

\title{
Cavity analysis on the robustness of random networks against targeted attacks:
influences of degree-degree correlations
}

\author{Yoshifumi Shiraki}
\email[Present address:]{shiraki@cs.brl.ntt.co.jp}
\altaffiliation{NTT Communication Science Laboratories, NTT Corporation,
3-1 Morinosato Wakamiya Atsugi-shi, Kanagawa, 243-0198, Japan}
\author{Yoshiyuki Kabashima}
\email[]{kaba@dis.titech.ac.jp}
\affiliation{Department of Computational Intelligence and Systems Science, Tokyo Institute of Technology, Yokohama 2268502, Japan}

\date{\today}

\begin{abstract}
We developed a scheme for evaluating the size of the largest connected subnetwork (giant component) in random networks and the percolation threshold when sites (nodes) and/or bonds (edges) are removed from the networks 
based on the cavity method of statistical mechanics of disordered systems. 
We apply our scheme particularly to random networks of bimodal degree distribution (two-peak networks), which have been proposed in earlier studies as robust networks against random failures of site and/or targeted (random degree-dependent) 
attacks on sites. 
Our analysis indicates that the correlations among degrees affect a network's robustness against targeted attacks 
on sites or bonds non-trivially depending on details of network configurations. 
\end{abstract}


\maketitle

\section{Introduction}
In the last decade, much research has been done on the regulation of connectivity of networks \cite{Strogatz2001,Kenah2007,Huang2006,Cohen2000,Dorogovtsev2008a}.
The Internet, electric power grids, and airline routes are examples of real-world networks, for which the connectivity 
is demanded
to resist various failures of nodes and/or links connecting the nodes. Preventing epidemics from spreading extensively in human/animal networks is another example of the regulation of network connectivity, for which, in contrast, reduction in the connection efficiency is required.

Erd\"{o}s-R\'{e}nyi-type random networks (graphs) \cite{Erdos1959,Erdos1960} and their derivatives provide research for the above-mentioned purposes with useful test beds. A basic random network is generated by connecting any pairs of sites (nodes) with a probability of $p$. The degree, which represents the number of bonds (edges) by which a site is connected directly to other sites, follows a Poissonian distribution, for which frequencies of large degrees that considerably deviate from the mean are practically negligible, 
in the case of the basic random network. However, real-world data indicate that there exist significantly many nodes of large degrees in the Internet, which is often modeled by a scale-free network, where the degree distribution is characterized by a power-law form $p(k) \propto k^{-\alpha}$ of an appropriate exponent $\alpha$.

Besides the degree distribution, there are several major feature quantities for characterizing network configurations. A na\"{i}ve generalization from the notion of the degree distribution may be the {\em degree-degree correlation} (or {\em degree correlation}), which represents the tendency of how likely sites of many bonds are to be connected to other sites of many or few bonds. The following two quantities are also widely used. One is the \textit{average path length}, which represents how many sites are at least necessary to follow a path between two randomly chosen sites. Randomly constructed networks typically have much shorter average path lengths than those for Cartesian lattices \cite{Dorogovtsev2008b}. The other is the so-called \textit{clustering coefficient}, which denotes the probability that a randomly chosen pair of sites, which come from the same particular site, is also connected. In our daily life, this implicates the tendency that two randomly chosen friends of a particular person are friends with each other as well \cite{Watts1998,Newman2000}.

Several earlier studies characterized the robustness of network connectivity using feature quantities, which originated in research on percolation phenomena \cite{Callway2000}. 
Cohen \textit{et al.} assessed the percolation threshold 
with respect to random removal of constituents on random networks,  
which was used for arguing the robustness of scale-free networks \cite{Cohen2000}. The method of generating function offers one of the most powerful techniques for such analyses. Based on this technique, Newman \textit{et al.} developed systematic schemes to compute the size of the largest subnetwork, which is occasionally termed the {\em giant component}, the percolation threshold, the average path length, etc. for random networks with arbitrary degree distributions \cite{Newman2001}. They clarified that scale-free networks are highly robust against random removal of sites and, in particular, scale-free networks of $\alpha > 3$ do not disintegrate unless all sites in networks break down. Newman also explored the effects of the degree correlation in random networks \cite{Newman2002}, and Golstev \textit{et al.} gave detailed consideration on how it affects the percolation threshold \cite{Goltsev2008}. They showed that so-called {\em assortative mixing} random networks exhibit higher robustness than those without degree correlations \cite{Vazquez2003a,Vazquez2003b}. The details are mentioned in the following sections. Some researchers investigated not only the robustness against random failures but also the effects of 
{\em targeted attacks}, 
the terminology of which is used for referring to {random degree-dependent attacks} throughout this paper, 
on sites on random networks \cite{Cohen2001,Kurant2007,Sole2008} and designed robust networks against various failures by optimizing the network configurations \cite{Tanizawa2005,Moreira2009}. Valente \textit{et al.} \cite{Valente2004} and Paul \textit{et al.} \cite{Paul2004} showed that random networks characterized by degree distributions of two or three peaks are most robust against targeted attacks and/or random failures.

Along this research direction, we mainly explore robustness of networks
against {\em targeted bond attacks}. 
More precisely, we particularly examine random networks of bimodal degree distributions (two-peak networks)
which are the most robust  
against 
site attacks.  
For this, we will develop 
an analytical scheme based on the {\em cavity method}, which was originally developed in statistical mechanics of disordered 
systems \cite{Mezard1987,Mezard2001}. 

For analyzing properties of large random networks, 
the cavity method approximately utilizes a message passing algorithm
that exactly holds in the Bethe lattices
under the assumption that the approximate treatment  
yields exact results for the random networks in the infinite system limit. 
Study on the Bethe lattices has a long history as one of few analytically 
solvable examples of percolation problems \cite{Stauffer1992}, and 
the equivalence between the infinitely large random networks and the Bethe lattices
has been shown for many examples of disordered systems \cite{Mezard2001}. 
In this sense, our approach can be considered as an extension of 
such existing studies to advanced settings in which 
various combinations of failure/attacks in conjunction with 
degree-degree correlations are taken into account.  

As shown later, our approach reproduces results identical to those
obtained by the {\em generating function method } \cite{Callway2000,Newman2001}
in various cases. Although the methodological relationship has not been fully clarified yet, 
this implies that the two schemes are potentially equivalent. 
However, the cavity method relies more highly on physical intuition and, therefore, 
may be easier to understand the physical meanings of relevant variables/equations 
that come out in the analysis.

This paper is organized as follows. In the next section, we briefly review several notions and known 
results concerning random networks which were 
necessary for our research. In particular, concerning the robustness of the connectivity, the optimality of the two-peak model, which we focus on later, is mentioned in some detail. In Sec. III, we 
develop a scheme for assessing feature quantities of percolation based on the cavity method. In Sec. IV, we discuss the use of our scheme for advanced problems of the two-peak model, in conjunction with validation by numerical experiments. The final section is devoted to the summary.

\section{Short Summary of Random Networks}
In this section, we review several notions and known results concerning random networks. The models that we focus on are also introduced. 

\subsection{Random Networks and Degree Distribution}
In general, a random network is composed of a collection of sites (nodes/points) and bonds (edges/links) that randomly connect pairs of sites. The number of bonds connected to a particular site is termed the {\em degree} $k$ of the site. 

{\em Degree distribution} $p(k)$, which denotes the frequency of sites that have degree $k$ in a network, is widely used for characterizing an ensemble of random networks. Conversely, it is often required to generate random networks following a given specific degree distribution $p(k)$. Let us denote $d_i$ as the degree of index $i$($=1,2,\dots,N$), where $N$ is the number of sites in the network. For $k=1,2,\dots,$ we set $d_i=k$ for $Np(k)$ indices of $i=1,2,\dots,N$. A practical scheme for the generation is basically as follows \cite{Steger1999}:
\begin{enumerate}
\item
[(P)] Make a set of indices $U$ to which each index $i$ attends $d_i$ times. Accordingly, we iterate (C1)--(C3).
\item
[(C1)] Randomly choose a pair of two different elements from $U$. We denote the indices of the two elements as $i$ and $j$.
\item
[(C2)] If $i \neq j$ and the pair of $i$ and $j$ has not been chosen up to that moment, make a bond between $i$ and $j$ and remove the two elements from $U$. Otherwise, we return them back to $U$.
\item
[(C3)] If $U$ becomes empty, finish the iteration. Otherwise, if it is not possible to make more bonds by (C1) and (C2), return to (P).
\end{enumerate}

Related to $p(k)$, one sometimes has to deal with the probability that one terminal of a randomly chosen bond has a degree $k$, which is assessed as 
\begin{align}
r_k&=\frac{kp(k)}{\sum _l l p(l)}\notag\\
&=\frac{kp(k)}{\langle k \rangle}
\end{align}
with use of the Bayes formula. Here, $\langle k \rangle=\sum_{l} l p(l)$ denotes the average of degrees on 
the random network. 

\subsection{Degree-Degree Correlations}
In addition to the frequencies, one can also take correlations of the degrees into account for characterizing properties of random networks by using the {\em joint degree-degree distribution} $r(k,l)$, which denotes a joint probability that two sites directly connected by a randomly chosen bond in a given network have degrees $k$ and $l$. We focus on {\em homogeneously random} networks, where $r(k,l)=r(l,k)$ holds for arbitrary pairs of $k$ and $l$. 

$r(k,l)$ has more information than $p(k)$ or $r_k$ in the sense that they are generally determined or reduced from $r(k,l)$ in such a way that 
\begin{eqnarray}
\sum_{l} r(k,l) = \frac{kp(k)}{\sum_{l} l p(l)}=r_k 
\end{eqnarray} 
holds for $\forall{k}$. Moreover, $r(k,l)$ is used for characterizing the degree correlations in random networks. In general, one can assess a conditional probability when one terminal site of a randomly chosen {\em bond} has degree $l$, the other terminal has degree $k$ as
\begin{eqnarray}
r_{k l}=\frac{r(k,l)}{r_l}=\frac{\left \langle k \right \rangle r(k,l)}{l p(l)}
\label{bond_conditioned}
\end{eqnarray}
from $r(k,l)$. Random networks, which satisfy
\begin{eqnarray}
r_{l k} = r_{l} \ \forall {k,l}
\label{no_correlation},
\end{eqnarray}
are generally regarded as {\em uncorrelated} or referred to as those of {\em no degree correlations}. 

Let us consider a situation in which we generate random networks following a given degree distribution $p(k)$. Even if $p(k)$ is specified, one can still control the degree correlations by designing conditional probabilities $r_{kl}$. The degree of freedom for such a design is assessed as
\begin{align}
F&=k_{\rm max}^2-\frac{k_{\rm max}(k_{\rm max}-1)}{2}-k_{\rm max}\notag \\
&=\frac{1}{2}\big( k_{\rm max}^2-k_{\rm max} \big), 
\end{align}
where $k_{\rm max}$ is the maximum degree in the network. This is because $k_{\rm max}^2$ positive values are assigned to $r_{kl}$ to satisfy $r_{kl}r_l=r_{lk}r_{k}$ for $\forall{k,l}$ and $\sum_l r_{kl}r_l=r_k$ for $\forall{k}$, which yields $k_{\rm max}(k_{\rm max}-1)/2$ and $k_{\rm max}$ constraints, respectively. 

The network generation scheme mentioned in the preceding subsection is not expected to produce any degree correlations. 
Many methods have been proposed for creating non-trivial correlations in random networks \cite{Newman2002,Maslov2002,Xulvi2004}.  
The scheme proposed in \cite{Newman2002}, 
which we employed for numerical experiments shown in the later sections,
generates random graphs so as to make the joint distribution 
of the {\em remaining degrees} $j$ and $k$ of two terminal sites of a randomly chosen bond
fit in a desired value $e_{jk}$.
The procedure of this scheme is summarized as follows: 
\begin{itemize}
\item[(I)]
For a given degree distribution $p(k)$, generate a basic graph by the algorithm shown in the preceding subsection. 
Accordingly, we iterate (D1)-(D3) sufficiently many times.  
\item[(D1)]
Choose two bonds denoted by pairs the terminal sites, $(v_1, w_1)$ and $(v_2, w_2)$,  randomly. 
\item[(D2)]
Evaluate remaining degrees $(j_1, k_1)$ and $(j_2, k_2)$ of the above two sites pairs. 
\item[(D3)]
Replace the bonds with two new site pairs
$(v_1, v_2)$ and $(w_1, w_2)$ with probability min$[1, (e_{j_1j_2}e_{k_1k_2})/(e_{j_1k_1}e_{j_2k_2})]$, 
\end{itemize}
where ${\rm min}(x,y)$ denotes the smaller value of $x$ and $y$.  
For macroscopically quantifying the degree-degree correlations, 
a measure 
\begin{equation}
R=\frac{1}{\sigma _q ^2}\sum _{jk}jk\big( e_{jk}-q_jq_k\big) \notag, 
\end{equation}
where
\begin{align}
\sigma _q&=\sum _k k^2q_k -\Big( \sum _k kq_k\Big) ^2\notag \\
q_k&=\frac{(k+1)p(k+1)}{\sum _j jp(j)}\notag, 
\end{align}
is often used \cite{Newman2002}.  
If $R$ of the above equation is zero, the random network is entirely random. On the other hand, positive (negative) $R$ indicates a tendency that the higher/lower-degree sites are more likely to be attached to other higher/lower-degree sites, the property of which is sometimes referred to as {\em assortative (disassortative) mixing}.

\subsection{Connectivity and Robustness}
We mainly explore the efficiency of connectivity of networks.
For this, let us define several notions concerning connectivity. 

We refer to two sites $i$ and $j$ as {\em connected} if and only if one can connect $i$ and $j$ by at least one sequence of bonds between them. Otherwise, the two sites are considered as {\em disconnected}. Given a network, a subnetwork in which any pairs of sites are connected is termed a {\em connected subnetwork}. In particular, a connected subnetwork that contains a majority of sites of the original network is often referred to as the {\em giant component}. 

When some sites and/or bonds are removed from the original network, the size of the giant component is reduced. Such removal can be regarded as corresponding to accidental/voluntary breakdown of machinery/connections in real-world networks. For theoretically assessing the robustness of network connectivity against such problems, we consider the following two sources of defects, {\em random failure}, which is modeled by statistically independent random removal of sites and/or bonds with certain probabilities and {\em targeted attack}, which is also dealt with as random removal of sites and/or bonds but only applied to sites whose degrees are larger than a certain threshold value. 

Let us denote $W_i$ as a probability that a site $i$ does \textit{not} belong to, or equivalently, is {\em not} connected to a giant component when the networks are randomly constructed. The size of the giant component is characterized by 
\begin{equation}
S=N^{-1}\sum _i(1-W_i). 
\label{giant_component}
\end{equation}
When the networks are affected by the above defect sources (failures/attacks), the typical value of $S$ is reduced as the defects strengthen. The critical value at which $S$ vanishes is termed the {\em percolation threshold}, by which we measure the robustness of network connectivity. 

\subsection{Two-Peak Random Networks}
Before closing this preparatory section, we mention two results about network robustness. 

Earlier independent studies \cite{Valente2004} and \cite{Paul2004} reported that networks characterized by bimodal degree distributions, which are occasionally referred to as {\em two-peak} networks, are most robust against random failures \textit{or} attacks to sites
among random network ensembles that are characterized only by the degree distributions
\footnote{Note that there is a slight difference between the two studies. Paul {\em et al.} \cite{Paul2004} emphasized that the most robust networks are achieved by {\em three-peak} distributions when the networks suffer from both failures \textit{and} attacks, while optimal networks against either failures {\em or} attacks are generally characterized by {\em two-peak} distributions. }.
Although both models are classified into the same category of two-peak networks, there is a difference between how the larger degree scales as the network size tends to infinity. 

Valente \textit{et al.} \cite{Valente2004} explored conditions of optimal network configurations under the constraint that network degrees are distributed in a bounded range $k \in [k_{\rm min},k_{\rm max}]$ ($0 \le k_{\rm min} \le k_{\rm max}$) keeping the average degree $\langle k \rangle$ a finite constant. They found that the optimal degree distribution against random failures is of the form that only $p(k_1)$ and $p(k_2)$ are finite while all other $p(k)$'s vanish, where $k_1=k_{\rm min}$ and $(k_{\rm min} \le ) k_2 (\le k_{\rm max})$ is determined in such a way that $\langle k^2 \rangle$ is maximized while keeping $\langle k \rangle$ constant. 

On the other hand, Paul {\em et al.} \cite{Paul2004} presented another type of two-peak network, which is robust against both random failures and targeted attacks. They assume that smaller-degree sites ($k_1$) are dominant in random networks and a very small number of the larger-degree ($k_2$) sites, which they call \textquotedblleft hub-nodes \textquotedblright, is inserted. 
Under the constraint that the number of sites $N$ and the average degree $\left \langle k \right \rangle$ are fixed, 
their finding indicates that the robustness is maximized when the following conditions hold \cite{Tanizawa2006}:
\begin{align}
k_1&\approx \langle k \rangle \text{,}  \notag \\
k_2&\approx \sqrt{\langle k \rangle N} \text{,} \notag \\
p(k_1)&=1-p(k_2) \notag \text{,} \\
p(k_2)&=\left( \left\{\frac{2\langle k \rangle ^2(\langle k \rangle
-1)^2}{2\langle k \rangle-1}\right\}^{2/3}/\langle k \rangle N \right)
^{\frac{3}{4}}. \notag
\end{align}
The networks become more robust for larger $N$. Note that $k_2$ in the above expression indicates the highest possible degree allowed in a random network with average $\langle k \rangle$ and size $N$ when no multiple bonds between identical pairs of sites are permitted \cite{Boguna2004}.

Hereafter, we refer to the former and latter as the {\em Valente-Sarkar-Stone (VSS) model} and the {\em Paul-Tanizawa-Havlin-Stanley (PTHS) model}, respectively. 

\section{Cavity Approach to Percolation Analysis}
In this section, we develop a scheme
for analyzing the robustness of random networks against various probabilistic attacks/failures based on the {\em cavity method} of statistical mechanics of disordered systems \cite{Mezard1987,Mezard2001}. 
Many existing studies on disordered systems indicate that the cavity method 
leads to the exact results in appropriate infinite system limits 
for various problems defined over
random networks that are characterized only by a degree distribution \cite{Mezard2001}
even when the support of the distribution is not bounded
\cite{Guerra2004,Talagrand2003,Giuraniuc2005}. 
In addition, recent studies imply that the coverage of this method 
can be extended to cases in which non-trivial degree-degree correlations exist
\cite{Coolen2008,Coolen2009,Coolen2010}.
Therefore, we expect that our analysis shown below provides
the exact results as well, the mathematically rigorous proof of which 
is unfortunately difficult. 
Instead, the validity of our scheme is shown by reproduction of known results, 
which have been obtained using other schemes in earlier studies, 
and comparison with numerical experiments. 

\subsection{Tree Approximation and Cavity Fields}
Let us develop an analytical scheme for evaluating the size of the giant component (Eq. (\ref{giant_component})) for assessing the network robustness. To do this, we first focus on a general property of randomly constructed networks, 
which is that lengths of closed paths between 
randomly chosen two sites (cycles) typically increase $O(\ln N)$ as the size of networks $N$ tends to infinity 
as long as the variance of the degree distribution is finite~\cite{Bollobas1985}. 
This property is presumably unchanged in most cases even when the degree-degree correlations 
are introduced as such correlations decrease exponentially fast with respect to 
lengths of paths except for extreme cases.  
This naturally leads us to approximately evaluate Eq. (\ref{giant_component}) 
by handling the local network structure of each site as 
that of {\em trees}, which are free from cycles.

To explain this, we use the notation $\partial i$ to denote a set of site indices $l$ that are directly connected to a site $i$. 
We also introduce an {\em indicator} $c_{l\to i}=0$ or $1$ in the \textit{i-cavity system} that 
is defined by removing site $i$ from the original system; 
$c_{l\to i}=0$ and $1$ indicate that $l$ does and does \textit{not} belong to a giant component 
of the \textit{i-cavity system}, respectively. 
Such variables are generally termed \textit{cavity fields} \cite{Mezard1987,Mezard2001}. A general and distinctive feature of trees is that all sites $l \in \partial i$ are completely disconnected from one another by removal of $i$. This means that when $i$ is inserted in the $i$-cavity system, while keeping the connection to a site $j \in \partial i$, $\left \langle i j \right \rangle $, is removed, 
the indicator that $i$ does not belong to the giant component is assessed as $\prod_{l \in \partial i \backslash j} c_{l \to i}$, where $X \backslash j$ denotes removal of $j$ from a set $X$. However, $i$ and $j$ are connected (a pair) through only the bond of $ \left \langle i j \right \rangle$ when the original system is a tree, this represents nothing but the cavity field of $i$ in the $j$-cavity system (see the left figure of Fig.1). This yields the equation
\begin{eqnarray}
c_{i \to j} = \prod_{l \in \partial i \backslash j} c_{l \to i}. \label{UC}
\end{eqnarray}
By moving entirely over the network, this equation can determine all the cavity fields $c_{l \to i}$ given appropriate initial and boundary conditions. After determining the cavity fields, 
the indicator that $i$ does not belong to a giant component of the original system, $o_i$, 
can be evaluated as
\begin{eqnarray}
o_i=\prod _{l\in \partial i}c_{l\to i}
\label{MP}
\end{eqnarray}
by taking the influences from all neighboring sites into account (the right figure of Fig.1).

\begin{figure*}[t]
\includegraphics{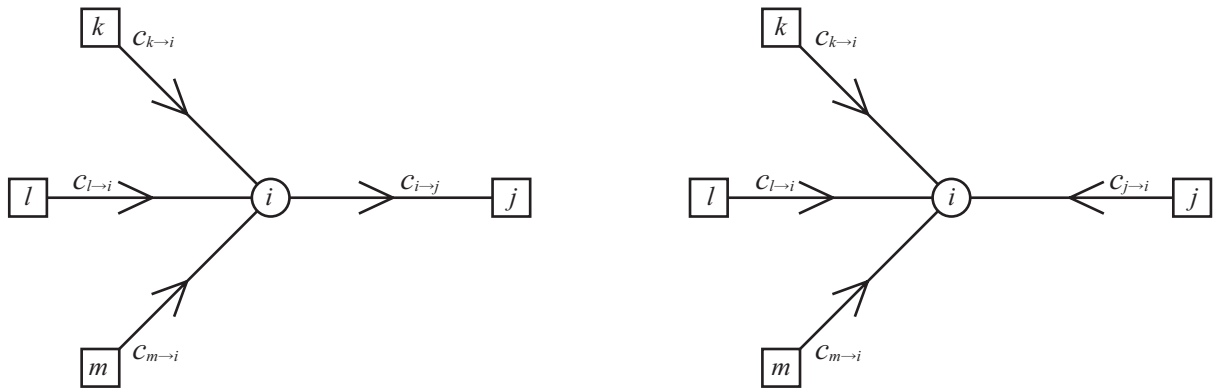}
\caption{Schematic diagram of computation in cavity method. The squares $k$, $l$, $m$, and $j$ denote sites in the $i$-cavity system. The left figure corresponds to Eq. (\ref{UC}). $c_{k \to i}$, $c_{l\to i}$, and $c_{m \to i}$ represent the cavity fields passed to $i$ when $i$ is inserted in the $i$-cavity system. By keeping the bond between $i$ and $j$ out, they compose the cavity field of $i$ in the $j$-cavity system, $c_{i \to j}$. After determining all cavity fields coming into $i$, $o_i$ is assessed, following Eq. (\ref{MP}) (the right figure). 
\label{cavity_method}}
\end{figure*}

\subsection{Macroscopic Description}
The following issue 
plays a key role for furthering our analysis. 
Let us classify sites in a graph by degree $k$, and denote $u_k$ as 
the frequency that the binary cavity fields $c_{l \to i}$ of sites $i$ of degree $k$ take the value of $1$. 
Namely, $u_k$ is defined as $u_k=(\sum_{i} \delta_{|\partial i|, k} k )^{-1} \sum_{i} 
\left (\delta_{|\partial i|, k} \sum_{l \in \partial i}
c_{l \to i} \right )$, where $|\partial i|$ stands for the degree of the site $i$ and 
$\delta_{x,y}=1$ if $x=y$ and $0$, otherwise. 
As $N \to \infty$, we can expect that $u_k$ typically converges to 
its average with respect to generation of graphs.
Such property is generally referred to as {\em self-averaging} \cite{Mezard1987}. 
In the current problem, the probability that a site neighboring a site of degree $k$ has a degree $m$ is denoted as $r_{mk}$.
Under the tree approximation, the influences of the cavity fields from the neighboring sites
can be averaged independently with respect to the graph generation. 
These mean that the algorithm of Eq. (\ref{MP}) can be macroscopically described as
{
\begin{align}
u_k&=\sum _{\mathbf{l}}\frac{(k-1)!}{l_1!l_2! \dotsm}\left(
r_{1 k}u_{1}\right) ^{l_1}\left( r_{2k}u_{2}\right) ^{l_2}\dotsm
\notag \\
&=\left( \sum _m r_{mk}u_m\right)^{k-1}, \label{macro_cavity_eq}
\end{align}
where $\sum _{\mathbf{l}}$ represents $\sum _{l_1}\sum _{l_2}\dotsm$ under the restriction of $\sum_t l_t=k-1$.
}

We assume that effects of random failures and/or targeted attacks can be taken into account by random removal of sites and/or bonds. For this, we denote $s_m$ and $b_{mk}$ as removal probabilities of sites of degree $m$ and bonds connecting two sites of degree $m$ and $k$, respectively. Suppose that under such an environment of random removal, a site $i$ is not connected to the giant component through the bond $\left \langle l i \right \rangle$ from a neighboring site $l \in \partial i \backslash j$ when $i$ is inserted into the $i$-cavity system. For this, at least one of the following 
{
three holds}:
\begin{enumerate}[(1)]
\item
$l$ is not connected to the giant component.
\item
$l$ is removed. 
\item
The bond $\left \langle l i \right \rangle $ is removed. 
\end{enumerate}
This indicates that $u_m$ in the right hand side in Eq. (\ref{macro_cavity_eq}) should be replaced by $1-(1-s_m)(1-b_{mk})(1-u_m)$ assuming that the degrees of sites $j$ and $i$ are $m$ and $k$, respectively. Substituting this into Eq. (\ref{macro_cavity_eq}) finally yields a set of equations for determining the frequencies of the cavity fields
\begin{equation}
u_k=\Big( 1-\sum _m r_{mk}(1-s_m)(1-b_{mk})(1-u_m)\Big) ^{k-1}
\label{eq:udilute}
\end{equation}
in a self-consistent manner. 

Let us denote $w_k$ as the probability that a site of degree $k$ is not connected to the giant component, 
which means that $W_i=w_k$ for a site $i$ of degree $k$. 
By averaging $(\sum_{i} \delta_{|\partial i|, k})^{-1} \sum_{i} \delta_{|\partial i|,k} o_i$ with 
respect to the graph generation and random (possibly degree-correlated) removal of sites/bonds, 
this is assessed as 
\begin{equation}
w_k=\Big( 1-\sum _m r_{mk}(1-s_m)(1-b_{mk})(1-u_m)\Big)^k,
\label{eq:qdilute}\\
\end{equation}
after determining $u_k$ for $\forall{k}=k_{\rm min}, \ldots, k_{\rm max}$ by the above set of equations. 
Consequently, the size of the giant component is evaluated as
\begin{eqnarray}
S=\sum_{k}p(k)(1-s_k)(1-w_k).
\label{eq:Sdilute}
\end{eqnarray}
These expressions constitute the main result of the current study. 

\subsection{Validation by Reproduction of Known Results}
Equations (\ref{eq:udilute}) and (\ref{eq:qdilute}) generally hold for random networks characterized by a joint degree-degree distribution $r(m,k)$. However, since the way of derivation and the obtained expressions are somewhat different from those of known analyses, it is not clear how our scheme is related to earlier studies. To clarify the relation, we first reproduce known results on the giant components for three basic examples using our developed scheme. 

\subsubsection{Random Failures on Uncorrelated Random Networks}
We first focus on {\rm uncorrelated networks}, which are characterized by $r_{mk}=r_k$ for $\forall{m,k}$. When $b_{mk}$ does not depend on the degrees of sites ($b_{mk}=b_m$), we can rewrite Eq. (\ref{eq:udilute}) as follows
\begin{align}
u_k&=\Big[1-\sum _m r_{m}(1-s_m)(1-b_m)(1-u_m)\Big]^{k-1}\notag\\
&=(1-U)^{k-1} \ \ \forall \ k \label{eq:uun}
\end{align}
where $U=\sum _m r_{m}(1-s_m)(1-b_m)(1-u_m)$. This yields a self-consistent equation of $U$ as
\begin{equation}
U=\sum _m r_{m}f_m(1-(1-U)^{m-1}),\label{eq:Uun}
\end{equation}
where $f_m$ is a product of fractions of site and bond failures $(1-s_m)(1-b_m)$. This means that the size of a giant component $S$ on a random network without degree correlation can be expressed using $U$ as
\begin{equation}
S=\Big( 1-s\Big) \sum _k p(k) \Big[ 1-(1-U)^k\Big], \label{eq:Sun}
\end{equation}
where $p(k)$ is a degree distribution of the random network. This expression is identical to the one obtained in earlier studies (for example \cite{Callway2000}). 

Equation (\ref{eq:Sun}) can be used for assessing a percolation threshold of an uncorrelated random network. For random failures of sites and bonds on random networks, $f_m$ does not depend on degrees $m$ of sites; therefore,
\begin{equation}
f=\frac{U}{\sum _m r_{m}(1-(1-U)^{m-1})}
\end{equation}
holds. This expression means that the giant component disappears when $U$ vanishes. Therefore, we take a limit $U \to 0$ in the above expression and use l'H\^opital's rule, which yields the percolation threshold as 
\begin{align}
f_c&=\frac{1}{\sum _m r_m (m-1)}\notag\\
&=\cfrac{1}{\cfrac{\langle k^2\rangle}{\langle k\rangle}-1}. \label{eq:falpha}
\end{align}
This also agrees with the one obtained in earlier studies (for example \cite{Newman2001}).

\subsubsection{Targeted Attacks on Uncorrelated Networks}
For the second example, let us assess the size of a giant component and the percolation threshold of uncorrelated networks against targeted attacks. For this, we assume that 
\begin{eqnarray}
b_{mk}= \left \{
\begin{array}{ll}
b_m, & m > m^\prime, \cr
0, & m \le m^\prime, 
\end{array}
\right .
\end{eqnarray}
which implies that the attack is targeted to only larger-degree sites. Let us denote $f_a$ and $f_r$ as rates of targeted attacks and random failures, respectively. When $f_r=0$ or there is no attacks to sites and/or bonds of degree $m \le m^\prime$, we have an expression
{
\begin{equation}
f_a=\frac{U-\sum _{m}r_{m}\{ 1-(1-U)^{m-1}\}}{\sum
_Mr_M\{1-(1-U)^{M-1}\}} \label{eq:fa0}
\end{equation}
where
$\sum _{m}=\sum _{k_{\rm min}} ^{m^\prime}$ 
and  $\sum _M=\sum _{m^\prime +1} ^{k_{\rm max}}$. 
Taking a limit of $U \to 0$ yields the percolation threshold as 
\begin{equation}
f_{a_c}=\frac{1-\sum _{m} r_{m} (m-1)}{\sum _M r_{M}
(M-1)}. \label{eq:fa0alpha}
\end{equation}
}
There could be cases where random networks still have giant components if all larger-degree sites are removed or $f_a=0$. In such cases, the percolation threshold of random failures after targeted removals is evaluated as 
\begin{equation}
f_{r_c}=\frac{1}{\sum _{m} r_{m}(m-1)}.\label{eq:fralpha}
\end{equation}
To evaluate the size of the giant component of random networks of no degree correlations, we rewrite Eq. 
(\ref{eq:Sun}) to make $s$ in Eq. (\ref{eq:Sun}) depend on $k$ as $s_k$, which offers an expression
\begin{equation}
S=\sum _k p(k) \Big( 1-s_k\Big)\Big( 1-(1-U)^k\Big). \label{eq:Stun}
\end{equation}

\subsubsection{Degree-Correlated Networks}
For the final example for validating our scheme, we analyze random networks of degree correlations against random failures. Unfortunately, it is difficult to obtain analytical expressions of thresholds for degree-correlated networks 
in the case of correlated networks. However, one can still numerically evaluate the threshold by solving 
\begin{align}
w_k&=\Big( 1-\sum _m r_{mk}(1-s)(1-b)(1-u_m)\Big) ^k, \label{eq:w_l}\\
u_k&=\Big( 1-\sum _m r_{mk}(1-s)(1-b)(1-u_m)\Big) ^{k-1}, \label{eq:u_l}\\
S&=(1-s)\sum _l p(l)(1-w_k),\label{eq:gcc}
\end{align}
which is computationally feasible.

One can use Eqs. (\ref{eq:qdilute}) and (\ref{eq:udilute}) 
to assess the percolation threshold due to bond attacks as follows.
{
Let $f=(1-s)(1-b)$ and $y_k=1-u_k$. This and Eq. (\ref{eq:udilute}) yield an expression }
\begin{equation}
1-y_{k}=\Big( 1-f\sum _m r_{mk}y_m\Big) ^{k-1}.
\end{equation}
Near the percolation threshold, $y_k \ll 1$ holds, which makes it possible to expand the left hand side of the above equations
\begin{equation}
y_k=f\sum _m(k-1)r_{mk}y_m,
\end{equation}
and express them in a matrix form,
\begin{equation}
\mathbf{y}=\mathbf{A}f\mathbf{y}\label{eq:matrix},
\end{equation}
where
\begin{equation}
A_{ij}=(i-1)r_{ij}.\notag
\end{equation}
The percolation threshold is determined by the condition that Eq. (\ref{eq:matrix}) has a nontrivial solution. This is reduced to the eigenvalue analysis of the matrix $\mathbf{A}$, and, in general, the largest eigenvalue corresponds to the percolation threshold.

To confirm the validity of our analytical scheme, we compared the size of the giant component predicted using Eq. (\ref{eq:gcc}) with the results of numerical experiments. The experiments were carried out for the two-peak networks of $k_1=4$ and $k_2=200$, which were designed to have an average degree of $4.4702$. The network size was set to $N=10000$. The configuration was obtained by optimizing networks based on the procedure proposed by Valente 
{\em et al.} \cite{Valente2004} under restrictions of $k_{{\rm min}}=4$, $k_{{\rm max}}=200$, and $\langle k \rangle=4.4702$. The identical configuration was also yielded using an optimization scheme proposed by Paul 
{\em et al.} \cite{Paul2004} for {\em random failures} under constraints of $k_1=4$ and $N=10000$. In addition to the optimized two-peak random network, we also performed experiments for non-optimal two-peak networks of $k_1=4$, $k_2=10$, $\langle k \rangle=4.4702$, and $N=10000$ for comparison. 

For two-peak networks, the joint distribution of the remaining degrees, $e_{jk}$, is uniquely determined 
when the network correlation $R$ is given in conjunction with degree distribution $p(k)$. 
Therefore, we controlled $R$ employing the scheme proposed by \cite{Newman2002}, which is summarized in Sec. II B., 
to the basic graphs constructed by the algorithm shown in Sec. II A. 
Analytically assessing the conditional degree distribution $r_{mk}$ for a given set of $e_{jk}$ is difficult. 
Therefore, we experimentally assessed $r_{mk}$ from the resulting networks and applied them to Eqs. (\ref{eq:w_l}) 
and (\ref{eq:u_l}) for evaluating the theoretical prediction of $S$. The values of the theoretical prediction numerically accord with those offered by an earlier study \cite{Goltsev2008} in high accuracy, although their methodological relation has not been clarified yet.
\begin{figure}
\includegraphics{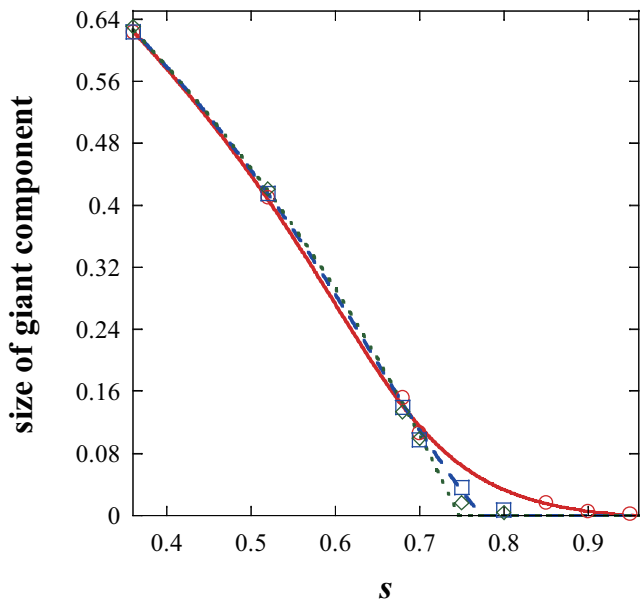}
\caption{(Color online) Sizes of giant component assessed using Eqs. (\ref{eq:udilute}) and (\ref{eq:qdilute}) for two-peak random networks that suffer from random failures. $s$ is a probability of site removal. The solid red curve shows the theoretical prediction for the optimized two-peak random network ($k_1=4$, $k_2=200$) of $R\approx 0$ (uncorrelated, solid red curve). Those for $k_1=4$ and $k_2=10$ of $R\approx -0.1$ (short-dashed green curve) and $R\approx 0.2$ (long-dashed blue curve) are also shown for comparison. The markers represent the results of numerical simulations, which are in good agreement with the theoretical predictions.
\label{rand}}
\end{figure}

Figure \ref{rand} shows the size of the giant component versus random failures of two-peak random networks optimized according to earlier studies and others to compare with the optimized ones. This indicates that networks become more robust as the degree correlation $R$ is increased, which has been pointed out by Newman \cite{Newman2002}.

\section{Analysis of Advanced Settings: Correlated Attacks for Correlated two-peak Networks}
In this section, we show 
the utility of the developed scheme by application
to advanced settings, which, as far as we know, have never been explored in earlier studies. More precisely, we consider problems of targeted removal of sites or bonds in two-peak random networks with degree correlations. A distinctive advantage of our scheme is the direct applicability to cases of correlated attacks and/or correlated networks, which was somewhat technically difficult in earlier studies. We use this advantage to examine how the degree-degree correlations affect the network robustness of two-peak random networks, which have been shown as optimal against both random failures and targeted attacks in the case with no degree-degree correlations. 

\subsection{Unified Treatment of Degree-Correlated Defects }
With regards to advanced settings in our scheme, let us consider situations in which targeted sites or bonds are removed in correlated two-peak random networks. We introduce the notation $f_{mk}$ to denote $(1-s_k)(1-b_{mk})$. Using this, Eq. (\ref{eq:udilute}) is expressed as 
\begin{equation}
u_l=\Big( 1-\sum _m r_{ml}f_{ml}(1-u_m)\Big) ^{l-1}, 
\end{equation}
which yields
\begin{equation}
\mathbf{By}=\mathbf{y}\label{eq:umatrix}
\end{equation}
for the matrix expression (\ref{eq:matrix}), where
\begin{equation}
B_{mk}=A_{mk}f_{mk}. 
\end{equation}
The percolation threshold again corresponds to a critical set of probabilities $f_{mk}$, which corresponds to a nontrivial solution of the above equation. This indicates that we can evaluate the percolation threshold by numerically solving
\begin{align}
\det[\mathbf{B}-\mathbf{E}]=0,\label{eq:dtB}
\end{align}
with respect to $f_{mk}$ and adopting one of the solutions.

\subsection{Degree-Correlated Site Attacks}
In general, there are multiple solutions that satisfy the above equations. However, we can still obtain analytical expressions of the percolation threshold by imposing some restrictions for the two-peak random networks. As one example, let us consider the cases where larger-degree {\em sites} are attacked preferentially under the constraint that the rate of site failures over all sites is fixed to a given value $s$. In such cases, we can generally set $f_a=f_{k_1k_2}=f_{k_2k_2}$ and $f_r=f_{k_1k_1}=f_{k_2k_1}$. These parameters are related with $s$ as
\begin{equation}
s=
\begin{cases}
(1-f_a) p(k_2), \ \ &0\leq s\leq p(k_2),\\
p(k_2)+(1-f_r)p(k_1), \ \ &s>p(k_2). 
\end{cases}
\label{s_two_possivilities}
\end{equation}
The first case of $0\leq s\leq p(k_2)$ in Eq. (\ref{s_two_possivilities}) implicates situations where only a portion of larger-degree sites suffers from the attacks, whereas the other portion and all smaller-degree sites are not damaged. In this case, Eq. (\ref{eq:dtB}) yields the percolation threshold as
\begin{equation}
s_c=
\left (1-\frac{A_{11}-1}{A_{22}(A_{11}-1) -A_{12}A_{21}} \right ) p(k_2). 
\end{equation}
The other case of $s>p(k_2)$ corresponds to situations where a giant component exists despite the fact that all larger-degree sites are removed, for which Eq. (\ref{eq:dtB}) offers the percolation threshold as 
\begin{equation}
s_c=p(k_2)+\left (1-\frac{1}{A_{11}} \right )p(k_1). 
\end{equation}

\begin{figure}[t]
\includegraphics{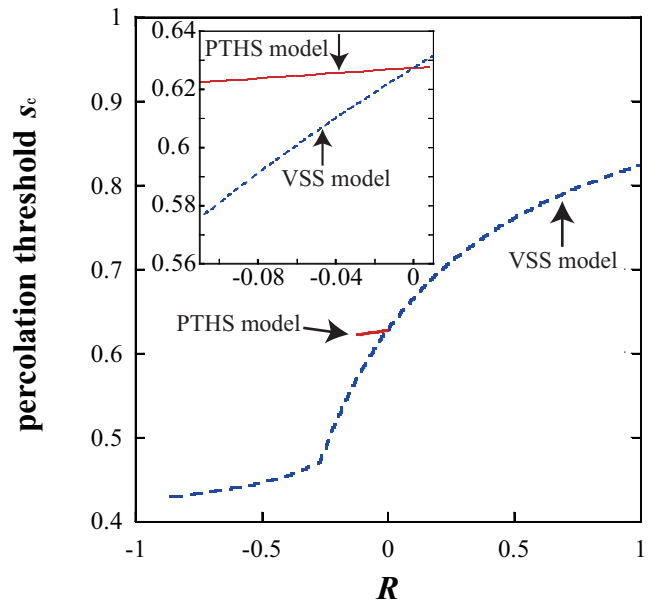}
\caption{(Color online) Percolation thresholds concerning site attacks $s_c$ versus degree correlation $R$. The solid (red) and dotted (blue) curves represent the percolation thresholds of the PTHS and VSS models, respectively. In the inset, the same data near $R=0$ are enlarged. The two curves intersect at $R=0$.
The reason why the data for the PTHS model is limited to a small region of $R$ is that the number of sites with $k_2=200$ degree
is so small that it is difficult to broadly change $R$. 
\label{phs_vs_r}}
\end{figure}

Figure \ref{phs_vs_r} shows the percolation thresholds assessed for the VSS and PTHS models, which are reviewed in Sec. II D. The VSS model has a network configuration with $k_1=4$, $k_2=5$, which is obtained using the optimization procedure of Valente {\em et al.} \cite{Valente2004} concerning {\em targeted attacks} on sites under the restriction of $k_{\rm min}=4$, $k_{\rm max}=200$, and $\langle k \rangle =4.4702$. On the other hand, the PTHS model has a network configuration with $k_1=4$, $k_2=200$, and $\langle k \rangle=4.4702$, which is offered with the optimization method proposed by Paul {\em et al.} \cite{Paul2004} under the restrictions of $k_1=4$ and $N=10000$.

In the two-peak models, Eq. (\ref{eq:fralpha}) guarantees that the value of the percolation threshold is identical between the two models for $R=0$. Earlier studies showed that the percolation thresholds concerning {\em random failures on sites and bonds} increase as $R$ increases for both models, implying that the assortiveness makes the network more robust against such failures. Figure \ref{phs_vs_r} indicates that this is the case for intentional site attacks as well. For all cases of the VSS model, the percolation threshold 
$s_c$
is determined by the second case in Eq. (\ref{s_two_possivilities}); namely, the giant component can be sustained even if all larger-degree sites are removed. However, for the 
PTHS model of sufficiently low $R (< -0.27)$, the giant component disappears only by removing a certain number of larger-degree sites. When $R$ is lower than zero, two sites of different degrees are more likely to be connected. This implies that larger-degree sites act like {\em glue} in forming a network of smaller-degree sites. Constituents of the network would be disconnected if the glue were removed. This may be the reason the giant component disappears by only removing a portion/certain number of larger-degree sites. 

\begin{figure}[t]
\includegraphics{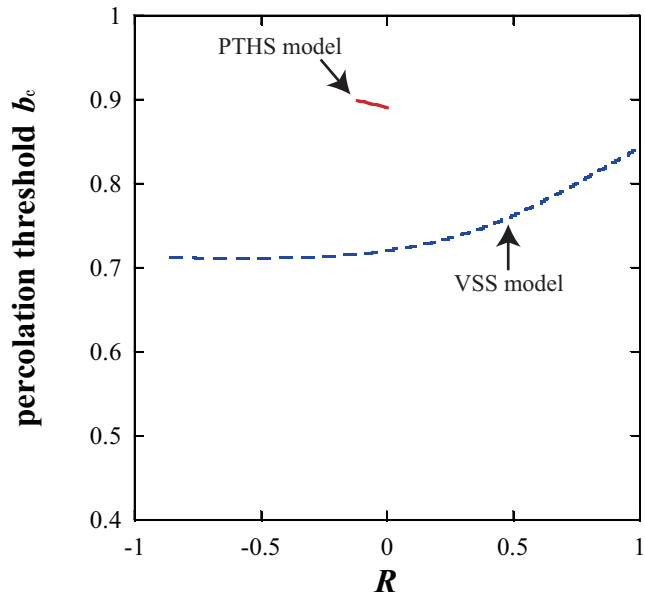}
\caption{(Color online) Percolation thresholds concerning bond attacks $b_c$ versus degree correlation $R$. The solid (red) and dotted (blue) curves represent the percolation thresholds of the PTHS and VSS models, respectively. The percolation threshold of the PTHS model decreases as $R$ increases.
The reason why the data for the PTHS model is limited to a small region of $R$ is identical to that of Fig. \ref{phs_vs_r}
\label{ph_vs_r}}
\end{figure}

Figure \ref{phs_vs_r} also indicates that the affect of $R$ is more significant in the VSS model than in the PTHS model. This might be understood as follows. In the PTHS model, the number of sites with degree $k_2$, $Np(k_2)$, grows as $O(N^{1/4})$. 
It is much smaller than that of bonds, which are directly linked to the sites, $k_2 N p(k_2) \sim O(N^{3/4})$. 
Even if $R$ is tuned, this prevents the sites from segregating from sites with degree $k_1$ and, therefore, the configuration of the network will not change drastically. This is in accordance with the profile of the percolation threshold in Fig. \ref{phs_vs_r}, which exhibits a weak dependence on $R$. In contrast, the numbers of sites and bonds for a larger degree $k_2$ 
are comparable  in the VSS model as $Np(k_2) \sim O(N)$ and $k_2 N p(k_2) \sim O(N)$ hold. This indicates that the larger-degree sites can form almost independent networks by themselves, separating from sites with degree $k_1$ when $R$ is set to a sufficiently large positive value. On the other hand, if $R$ is set to a negative value, connections of sites of the two different degrees are enhanced. These mean that the profile of the network is highly affected by the regulation of $R$, which may lead to a relatively strong dependence of the percolation threshold on $R$ in Fig. \ref{phs_vs_r}.

\begin{figure}[t]
\includegraphics{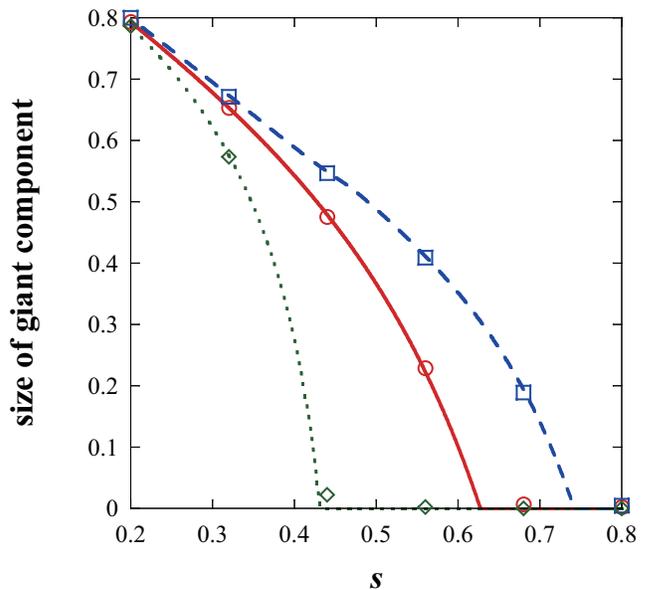}
\caption{(Color online) Sizes of giant component versus targeted site attacks. The sizes of the giant component are plotted for the PTHS model with $R\approx 0$ (analytical results: solid red line/numerical simulation: circles) and the VSS model with $R\approx 0.4$ (dashed blue line/squares) and $R\approx -0.8$ (dotted green line/diamonds). The network size for the numerical simulation is $N=10000$. The error bars are smaller than the markers.
\label{sitea}}
\end{figure}

\begin{figure}[t]
\includegraphics{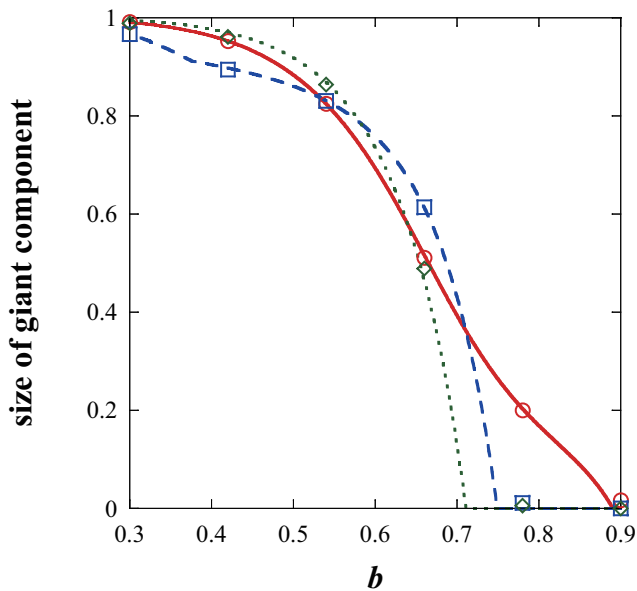}
\caption{(Color online) Sizes of giant components versus targeted bond attacks. The sizes of the giant component are plotted for the PTHS model with $R\approx 0$ (analytical results: solid red line/numerical simulation: circles) and the VSS model with $R\approx 0.4$ (dashed blue line/squares), and $R\approx -0.8$ (dotted green line/diamonds). The network size for the numerical simulation is $N=10000$ as well as the simulation in Fig. \ref{sitea}
The error bars are smaller than the markers. 
\label{bonda}}
\end{figure}

\subsection{Degree-Correlated Bond Attacks}
For another example, we consider a situation in which the attacks are preferentially launched on {\em bonds} connecting to larger-degree sites under the constraint that the rate of failures over all bonds is fixed to $b$. We set $f_a=f_{k_2k_2}$ and $f_r=f_{k_1k_1}=f_{k_2k_1}=f_{k_1k_2}$ as we did in the analysis on site attacks. This yields an expression 
\begin{align}
b&=\notag\\
&\begin{cases}
(1-f_a)r(k_2,k_2), \ \ \ \ \ \ \ \ \ \ \ \ \ \ \ \ \ \ 0\leq b\leq r(k_2,k_2),\\
r(k_2,k_2)+(1-f_r)(1-r(k_2,k_2)),\ b> r(k_2,k_2). 
\end{cases}
\label{b_two_possivilities}
\end{align}
The first case of $0\leq b \leq r(k_2,k_2)$ in Eq. (\ref{b_two_possivilities}) implies situations where only a fraction of bonds connecting two larger-degree sites suffers from attacks, whereas all the other bonds are not damaged. In this case, Eq. (\ref{eq:dtB}) yields the percolation threshold as
\begin{equation}
b_c=
\left ( \frac{(A_{11}-1)(A_{22}-1)-A_{12}A_{21}}{A_{22}(A_{11}-1)} \right ) r(k_2,k_2). 
\end{equation}
The other case of $b>r(k_2,k_2)$ corresponds to situations where a giant component exists even when all bonds connecting two larger-degree sites are removed, for which Eq. (\ref{eq:dtB}) offers the percolation threshold as 
\begin{equation}
b_c=r(k_2,k_2)+b_{r_c}[1-r(k_2,k_2)],
\end{equation}
where
\begin{equation}
b_{r_c}=\left ( \frac{2A_{12}A_{21}+A_{11}-\sqrt{A_{11} ^2+4A_{12}A_{21}}}{2A_{12}A_{21}} \right ).
\end{equation}

Figure \ref{ph_vs_r} indicates that the VSS model is not as robust as the PTHS model. This may be understood as follows. In bond attacks, bonds connecting two sites of a larger degree $k_2$ are removed preferentially. In the PTHS model, the number of the sites of degree $k_2$ is much less than that in the VSS model. This implies that the bonds that suffer from these attacks are relevant only to such a small number of sites in the PTHS model and, therefore, the damage on such bonds will not significantly deteriorate the total efficiency of the network connectivity compared to the case of the VSS model. Another distinctive feature is that the percolation threshold of the PTHS model decreases as $R$ increases. When $R$ is set to a negative value, the number of bonds that link two nodes of different degrees increases. This decreases the number of bonds linking two larger-degree sites, which will work as a factor of deterioration of the percolation threshold. However, at the same time, the increase in the number of bonds connecting two sites of different degrees will enhance the network connectivity. The dependence of the percolation threshold on $R$ implies that this effect overcomes the deterioration factor in the PTHS model. 

\subsection{Numerical Validation}
For justifying the results obtained in this section, we performed numerical simulations for 
the two types of two-peak-correlated random networks. 
The procedures for generating these random networks are the same as those mentioned in Sec. III C.

Figures \ref{sitea} and \ref{bonda} show the size of the giant component against targeted attacks on sites and bonds, respectively, for the PTHS and VSS models of $N=10000$. The configurations are identical to those mentioned in Sec. IV B. One can find that analytical results obtained with our scheme agree with those from numerical simulations with high accuracy. 

In these figures, changes in $R$ cause little notable difference as long as 
{
the situation}
is set far from the percolation threshold. However, the figures also 
indicate that the giant component generally decreases as $R$ increases. In addition, there is a tendency that a network with higher robustness has a smaller giant component at moderate number of removals.
This indicates that we should pay attention not only to the percolation threshold but also to the size of the giant component when we want to construct a robust network.

\section{Summary}
We have developed a scheme for analyzing random networks characterized by arbitrary degree distributions based on the cavity method of statistical mechanics of disordered systems. By approximately regarding the local structure of a given network as a tree, one can construct an approximation method for assessing the probability that a site does not belong to the giant component, which is the largest connected subnetwork of a random network. The most distinctive advantage of the scheme is the wide applicability against various types of attacks and failures that are intended for sites and/or bonds, which holds even in the presence of degree correlations. The validity and usability of the scheme have been shown by comparing it with known results and applying it to analysis of site/bond attacks in degree-correlated networks of bimodal degree distributions (two-peak networks) in conjunction with numerical justification. 

Promising future works include assessment of robustness against bond attacks for more realistic networks. Generalizing our approach to multi-bond and/or directed networks may also be of interest. 

\vspace*{2\baselineskip}

\section*{Acknowledgements}
The authors thank Toshihiro Tanizawa and Hisanao Takahashi for useful comments and discussions. This work was partially supported by Grants-in-Aid for Scientific Research on the Priority Areas ``Deepening and Expansion of Statistical Mechanical Informatics'' from the Ministry of Education, Culture, Sports, Science and Technology, Japan and KAKENHI No. 22300003 from JPSP.


%

\end{document}